\documentclass[prb,aps,twocolumn,amsmath,amssymb,showpacs]{revtex4}
\usepackage{graphicx}
\usepackage[english]{babel}
\usepackage{natbib}
\usepackage{amsfonts}
\usepackage{amssymb}
\usepackage{dcolumn}
\usepackage{amsmath}
\usepackage[ansinew]{inputenc}
\usepackage{amssymb}

\begin{document}
\title{Driven flux-line lattices in the presence of weak random columnar
    disorder: Finite-temperature behavior and dynamical melting of
    moving Bose glass}

\author{Y. Fily}
\author{E. Olive}
\author{J.C. Soret}
\affiliation{LEMA, UMR 6157, Universit\'e F. Rabelais-CNRS-CEA, Parc de Grandmont, 37200 Tours, France}

\begin{abstract}
We use 3D numerical simulations to explore the phase diagram of driven flux line lattices in presence of weak random columnar disorder at finite temperature and high driving force.
We show that the moving Bose glass phase exists in a large range of temperature, up to its melting into a moving vortex liquid.
It is also remarkably stable upon increasing velocity : the dynamical transition to the correlated moving glass expected at a critical velocity is not found at any velocity accessible to our simulations.
Furthermore, we show the existence of an effective static tin roof pinning potential in the direction transverse to motion, which originates from both the transverse periodicity of the moving lattice and the localization effect due to correlated disorder. Using a simple model of a single elastic line in such a periodic potential, we obtain a good description of the transverse field penetration at surfaces as a function of thickness in the moving Bose glass phase.
\end{abstract}

\maketitle

Periodic structures driven on a random substrate, such as vortex lattices in type II superconductors, exhibit a rich variety of phases controlled by the interplay between elasticity, disorder, temperature, and driving force \cite{Koshelev1994,Giamarchi1996,Balents1997,Simul1}. At large enough velocity, where the efficiency of the quenched disorder is reduced leading to dynamical reordering \cite{Simul2}, several moving glass phases were theoretically predicted \cite{Giamarchi1996,Balents1997}.
In particular, considering elastic deformations for weak disorder or large velocity in $d=3$, a topologically ordered phase is predicted \cite{Giamarchi1996,Simul3}, namely the moving Bragg glass (MBG).
%
The extension to correlated disorder led to the prediction of the moving Bose glass (MBoG) \cite{Chauve2000a}, characterized by the dynamical transverse Meissner effect (DTME), {\it i.e.} the tilt response to transverse field vanishes below a critical value.
At finite temperature, renormalization group calculations \cite{Chauve2000a} show a transition at a critical velocity $v_c$ from the MBoG to a very high velocity glassy phase in which the DTME vanishes, namely the correlated moving glass (CMG).
Clear evidence of the MBoG was found in numerical simulations at $T=0$ \cite{Olive2003}.
However, a complete theory of elastic medium at high velocity is lacking, and the stability of moving elastic phases such as MBG or MBoG at the thermodynamic limit is still debated.

In this paper, we perform 3D molecular dynamics simulations of superconductor vortices with weak random columnar pinning.
We show in details the existence of the MBoG at finite temperature. It appears so stable that the expected dynamical transition to the CMG is not found.
Furthermore, we find the existence of an effective pinning potential that is $z$ independent ($z$ being the direction of the columnar pins) and periodic in the direction transverse to motion.
Such effective pinning potential appears as an additional signature of the MBoG. Consequently, we extend to finite temperatures a model of a single elastic line into a tin roof potential \cite{Olive2003}, which captures the DTME property of MBoG and yields quantitative understanding of finite thickness effects.
Finally, the dynamical melting of MBoG is studied as temperature is increased.

Following Ref. \cite{Olive2003}, we model a stack of $N_z$ Josephson-coupled parallel superconducting planes of thickness $d$ with interlayer spacing $s$. Each layer in the $(x,y)$ plane contains $N_v$ pancake vortices interacting with $N_p$ columnar pins parallel to the $z$ direction. The overdamped equation of motion of a pancake $i$ at position $\bf r_i$ reads
\begin{equation}
\nonumber
\begin{split}
\eta \frac{d{\bf r_i}}{dt}=- & {\sum_{j \neq i}}\nabla_i U^{vv}(\rho_{ij},z_{ij})-{\sum_{p}}\nabla_i U^{vp}(\rho_{ip})  \\
 & +{\bf F}^L+{\bf F}^{tilt}(z)+{\bf F_i}^{th}(t)
\end{split}
\end{equation}
\noindent
where $\rho_{ij}$ and $z_{ij}$ are the components of ${\bf r_{ij}}={\bf r_i}-{\bf r_j}$ in cylindrical coordinates, $ \rho_{ip}$
 is the in-plane distance between the pancake $i$ and a pinning site in the same layer at ${\bf r_p}$, and $\nabla_i$ is the 2D gradient operator acting in the 
$(x,y)$ plane. $\eta$ is the viscosity coefficient, ${\bf F}^L=F^L{\bf \hat x}$ is the Lorentz driving force due to an applied current, ${\bf F_i}^{th}$ is the thermal noise with $\langle F_{i,\alpha}^{th} \rangle=0$ and $\langle F_{i,\alpha}^{th}(t) F_{j,\beta}^{th}(t') \rangle=2 \eta k_B T \delta_{ij} \delta_{\alpha \beta} \delta (t-t')$ where $\alpha,\beta=x,y$ and $k_B=1$ is the Boltzmann constant.
${\bf F}^{tilt}(z)$ is the surface force due to the field tilting away from the $z$ axis  in the $y$ direction. This force acts as a torque on each flux line, {\it i.e.} 
${\bf F}^{tilt}(z=0)=-{\bf F}^{tilt}(z=N_zs)=F^{tilt}{\bf \hat y}$ and ${\bf F}^{tilt}(z)={\bf 0}$ for pancakes in the bulk. The tilting force modulus is defined by
$F^{tilt}=\epsilon^2\phi_0H_y/4\pi=8\pi\epsilon^2\epsilon_0\lambda_{ab}^2H_y/\sqrt 3 a_0^2H_z$, where $\epsilon_0=(\phi_0/4\pi\lambda_{ab})^2$, 
$\lambda_{ab}$ is the in-plane magnetic penetration depth, $a_0$ is the average vortex distance, $H_y$ is the transverse field component, and $\epsilon$ is the 
anisotropy parameter. 
The intra-plane vortex-vortex repulsive interaction is given by a modified Bessel  function $U^{vv}(\rho_{ij})=2\epsilon_0dK_0(\rho_{ij}/\lambda_{ab} )$. 
The inter-plane attractive interaction between pancakes in adjacent layers of altitude $z$ and $(z+s)$ reads 
$U^{vv}(\rho_{ij},z_{ij}=s)=(2s\epsilon_0/\pi)[1+ln(\lambda_{ab}/s)][(\rho_{ij}/2r_g)^2-1]$ for $\rho_{ij}\le 2r_g$ 
and $U^{vv}(\rho_{ij},z_{ij}=s)=(2s\epsilon_0/\pi)[1+ln(\lambda_{ab}/s)][\rho_{ij}/r_g-2]$ otherwise ; in this expression $r_g=\xi _{ab}/\epsilon $,
where $\xi_{ab}$ is the in-plane coherence length. This pairwise interaction results from both electromagnetic and josephson coupling \cite{Ryu1992}.
Finally, the attractive pinning potential is given by $U^{vp}( \rho_{ip})=-\alpha A_pe^{-(\rho_{ip}/R_p)^2}$, where  
$A_p=(\epsilon _0d/2)ln[1+(R_p^2/2\xi_{ab}^2)]$ \cite{Blatter1994} and $\alpha$ is a tunable parameter.
We consider periodic boundary conditions of $(L_x, L_y)$ sizes in the $(x,y)$ plane while open boundaries are taken in the $z$ direction. Molecular dynamics simulation is used for $N_v$ vortex lines in a rectangular basic cell $(L_x,L_y)=(5, 6 \sqrt3/2) t \lambda_{ab}$, with $t=1,2$, and for a number of layers varying from $N_z=19$ to $N_z=1999$. All details about our method for computing the Bessel potential with periodic conditions can be found in Ref. \cite{Olive1998}.
The number of columnar pins is set to $N_p=N_v$. We consider the London limit $\kappa =\lambda_{ab} /\xi_{ab} =90$, with an average vortex distance 
$a_0=\lambda_{ab} $, and  $d=2.83\ 10^{-3}\lambda_{ab}$,  $s=8.33\ 10^{-3}\lambda_{ab}$, $R_p=0.22\ \lambda_{ab}$, $\epsilon =0.01$, $\eta=1$.
We choose the tunable pinning parameter $\alpha =1/25$ so that the maximum pinning force is $F_{max}^{vp}\sim F_0/5$ where $F_0=2\epsilon _0d/\lambda_{ab}$ is the unit force defined by the Bessel interaction.
All the parameters values are identical to our $T=0$ previous study \cite{Olive2003} so that a direct comparison is possible.

\begin{figure}[t]
\centering
\includegraphics[width=0.95\linewidth]{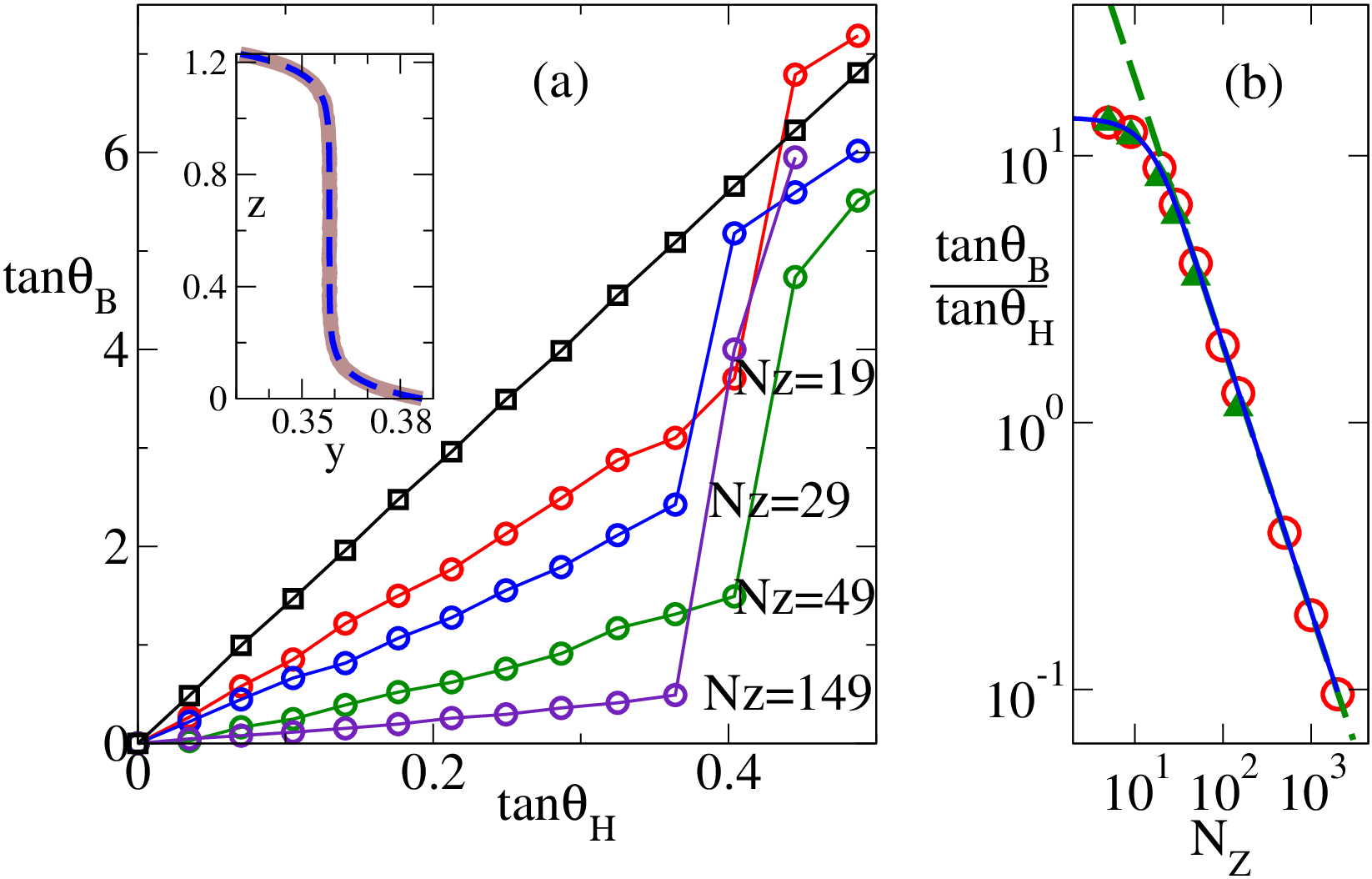}
\caption{ (color online)
(a) Average vortex line inclination versus field inclination at $T=10^{-4}$ for $v=10^{-2}$, $N_v=30$ and several thicknesses (circles) compared with disorder free linear response (squares). Insert : average shape of a line at low inclination
for $N_z=149$ and $T=10^{-10}$ (thick brown line), and $\sinh$ fit (thin dashed blue line).
(b) Slope $d (\tan \theta_B)/d (\tan \theta_H)$ at low angle versus $N_z$ for $T=10^{-10}$ (red circles) and $T=10^{-4}$ (green triangles).
The solid blue line is the $\tanh\left(\frac{L}{2 z_0}\right)/L$ fit, the dashed green line is the $N_z^{-1}$ dependence at large $N_z$.}
\label{TME}
\end{figure}

The driving force ${\bf F}^L$ is chosen high enough to obtain a fully elastic flow at $T=0$. 
At low temperature, the vortex flow remains elastic with no dislocations. The rough static channels observed at $T=0$ persist, except that they are broadened by thermal fluctuations.
Starting with such a lattice moving in the $x$ direction and the magnetic field $H$ along the $z$ axis, we slowly vary the $y$ component of $H$ ($H_z$ being fixed) in order to obtain the transverse induction response. Fig.\ \ref{TME}a shows the flux line inclination $\tan \theta_B=B_y/B_z$ averaged over time versus the magnetic field inclination $\tan \theta_H=H_y/H_z$ for several thicknesses $N_z$. The response is linear at low angle, with a finite slope which decreases when thickness is increased and eventually vanishes in the $N_z \rightarrow \infty$ limit, as shown in Fig.\ \ref{TME}b. Such finite slope is explained in the insert of Fig.\ \ref{TME}a : the lines are curved at their extremities by the tilt force while in the bulk they remain aligned with the $z$ axis, \textit{i.e.} the transverse field only penetrates the sample near its surfaces, resulting in the partial screening of $H_y$.
This supports that the finite response at low angle is a surface effect. Fig.\ \ref{TME}a also shows that at a critical transverse field, the line inclination experiences a jump associated with an angular depinning transition of the vortex lines.
Above this transition, the lines display a kink structure in the $yz$ planes (see Fig. 3 of Ref. \cite{Olive2003}).
$dz/dy$ is almost independent of the line for a given $y$ and periodic in the $y$ direction,
indicating that all the lines experience the same effective pinning landscape, which has the invariance of the columnar pins and the transverse periodicity of the channels.
These results, very similar to those obtained at $T=0$, strongly suggest the existence of the MBoG phase at finite temperature.
Since a finite transverse magnetic response at low angle is expected in the CMG \cite{Chauve2000a}, a careful study of finite size effects in the $z$ direction is crucial to discrimate MBoG from CMG.
In order to understand quantitatively the influence of thickness, we study a model introduced in Ref. \cite{Balents1995} and used in Ref. \cite{Olive2003} to describe the tilt of the lines at $T=0$, that we phenomenologically apply to quantities averaged over time in the finite temperature case. It is a mean field approach describing the angular response of the vortex lattice in terms of a single elastic line put in the effective pinning potential $V(y)$ discussed above. The energy $E$ of a line of length $L=N_z s$ is given by
\begin{eqnarray}
\nonumber
E(u)=\int_{0}^{L}dz\left (\frac{c}{2}\left (\frac{du}{dz}\right )^2+V(u)\right)+f\left(u(L)-u(0)\right)
\end{eqnarray} 
where $u(z)$ is the one-dimensional displacement field in the $y$ direction, $c=\epsilon^2\epsilon_0$ is the elastic constant and $f\propto H_y/H_z$ is a surface force. This expression of the energy doesn't take explicitly into account thermal fluctuations, however line elasticity and effective potential can depend on temperature.
Minimizing $E$ with respect to $u(z)$ while expanding $V$ quadratically near a minimum leads to the following solution $u(z)$ for a line~: 
$u(z)=\frac{z_0 f}{c}\sinh\left(\frac{z-L/2}{z_0}\right)/\cosh\left(\frac{L}{2 z_0}\right)$, 
where $z_0=\sqrt{c\left(\frac{d^2V}{du^2}\right)_{u=0}^{-1}}$ characterizes the thickness of the region where the transverse field penetrates the sample near the surface. The average line inclination is given by $\tan \theta_B =\left[u(L)-u(0)\right]/L=\frac{2 z_0 f}{cL} \tanh\left(\frac{L}{2 z_0}\right)$.
The above equation for $u(z)$ is observed to accurately fit the data, as shown in the insert of Fig.\ \ref{TME}a.
$z_0$ calculated from this fit is found to be independent of both $N_z$ and $\theta_H$, which is consistent with the observation that at low $\theta_H$ (\textit{i.e.} when $u(z)$ is small enough for the quadratic expansion to be accurate), $\tan \theta_B$ is a linear function of $\tan \theta_H$ (see Fig.\ \ref{TME}a) and $\tan \theta_B \propto \tanh\left(\frac{L}{2 z_0}\right)/L$ (see Fig.\ \ref{TME}b) which is verified up to large thicknesses compared with the penetration length. In the large thickness limit $L\gg z_0$, $\tanh\left(\frac{L}{2 z_0}\right) \sim 1$ and the inclination scales as $L^{-1}$, \textit{i.e.} as $N_z^{-1}$. From this we conclude to a true vanishing response to transverse field below a threshold in the infinite thickness limit, confirming that we are seeing MBoG and not CMG.
Finally, our results support the existence of the effective pinning potential at $T>0$ and reinforce the interest of the simple mean field model --- extended to finite temperatures provided that simulated quantities are averaged over thermal fluctuations --- as a tool to describe DTME, including finite size effects.
These behaviors have been observed in a wide range of velocities, from the appearance of the MBoG to the highest velocities we can simulate because of duration issues (over more than $4$ orders of magnitude at $T=10^{-7}$, over $3$ orders of magnitude at $T=10^{-4}$), \textit{i.e.} we don't see the expected dynamical transition to CMG \cite{Chauve2000a}.
Since the critical velocity $v_c$ at which the transition to CMG is expected behaves like $1/L_c\propto N_p \alpha^2$ where $L_c$ is the static Larkin length, a similar analysis has been performed for weaker pinning strengths. No evidence of CMG phase has been found in that case either.

\begin{figure}[h]
\centering
\includegraphics[width=0.67 \linewidth]{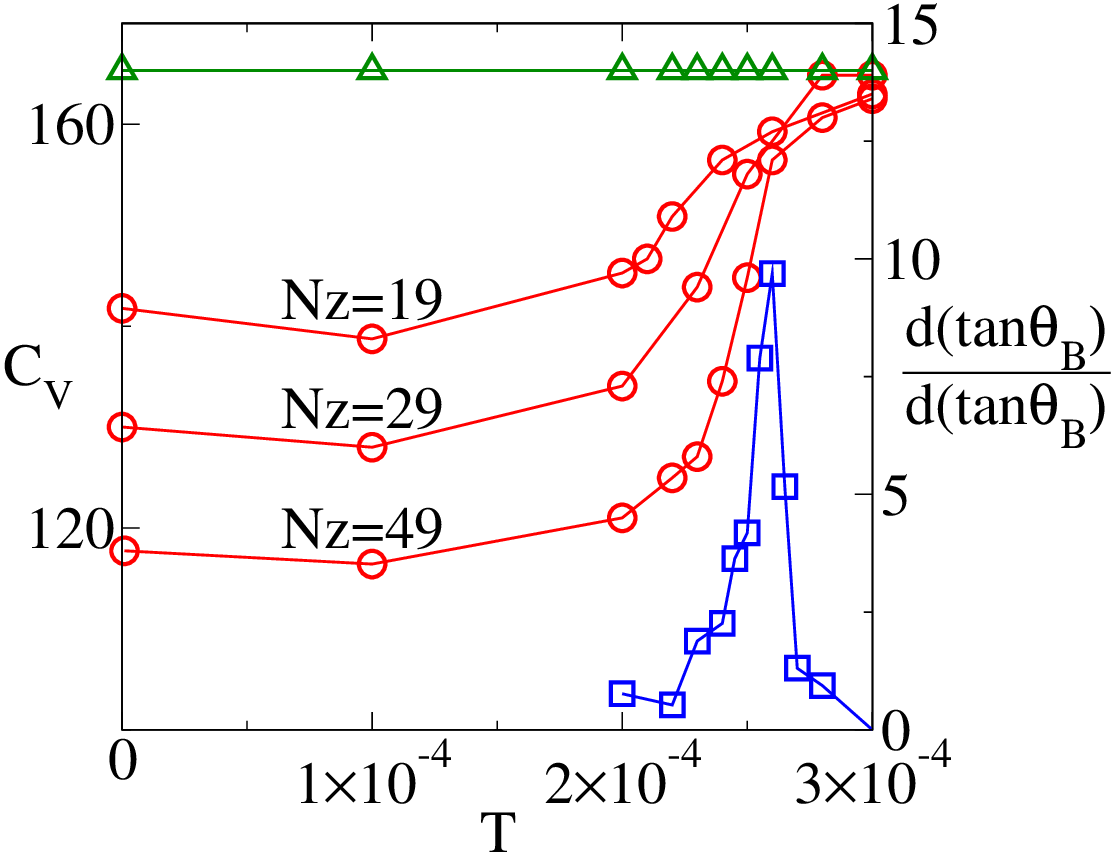}
\caption{ (color online)
$C_V=\frac{\langle E_t^2 \rangle-\langle E_t \rangle^2}{k_B T^2}$ with $k_B=1$ for $N_z=49$ (blue squares) and slope of the transverse induction response $d (\tan \theta_B)/d (\tan \theta_H)$ at low angle versus temperature for $N_z=19,29,49$ (red circles) compared with the pinning free case (green triangles), both for $N_V=30$ and $v=10^{-2}$.
}
\label{CvdB}
\end{figure}

We now fix the velocity ($v=10^{-2}$) and study the effect of temperature.
At zero tilt ($\theta_H=0^\circ$), we compute $C_V=(\langle E_t^2 \rangle-\langle E_t \rangle^2)/k_B T^2$ (Ref. 13) where $E_t$ is the total interaction energy ($C_V$ would be the specific heat at thermodynamic equilibrium). Together with $C_V$, we plot in Fig.\ \ref{CvdB} the slope $d (\tan \theta_B)/d (\tan \theta_H)$ of the linear region of Fig.\ \ref{TME}a at low angle. At low temperature $d (\tan \theta_B)/d (\tan \theta_H)$ is reduced compared with the pinning free case and decreases when thickness is increased, illustrating the DTME property as seen previously. It experiences a jump around $T=2.5 \times 10^{-4}$ while the thickness dependence vanishes, what we interpret as the loss of DTME.
 Concomitantly, $C_V$ exhibits a sharp peak suggesting a dynamical phase transition.
\begin{figure}[t]
\centering
\includegraphics[width= \linewidth]{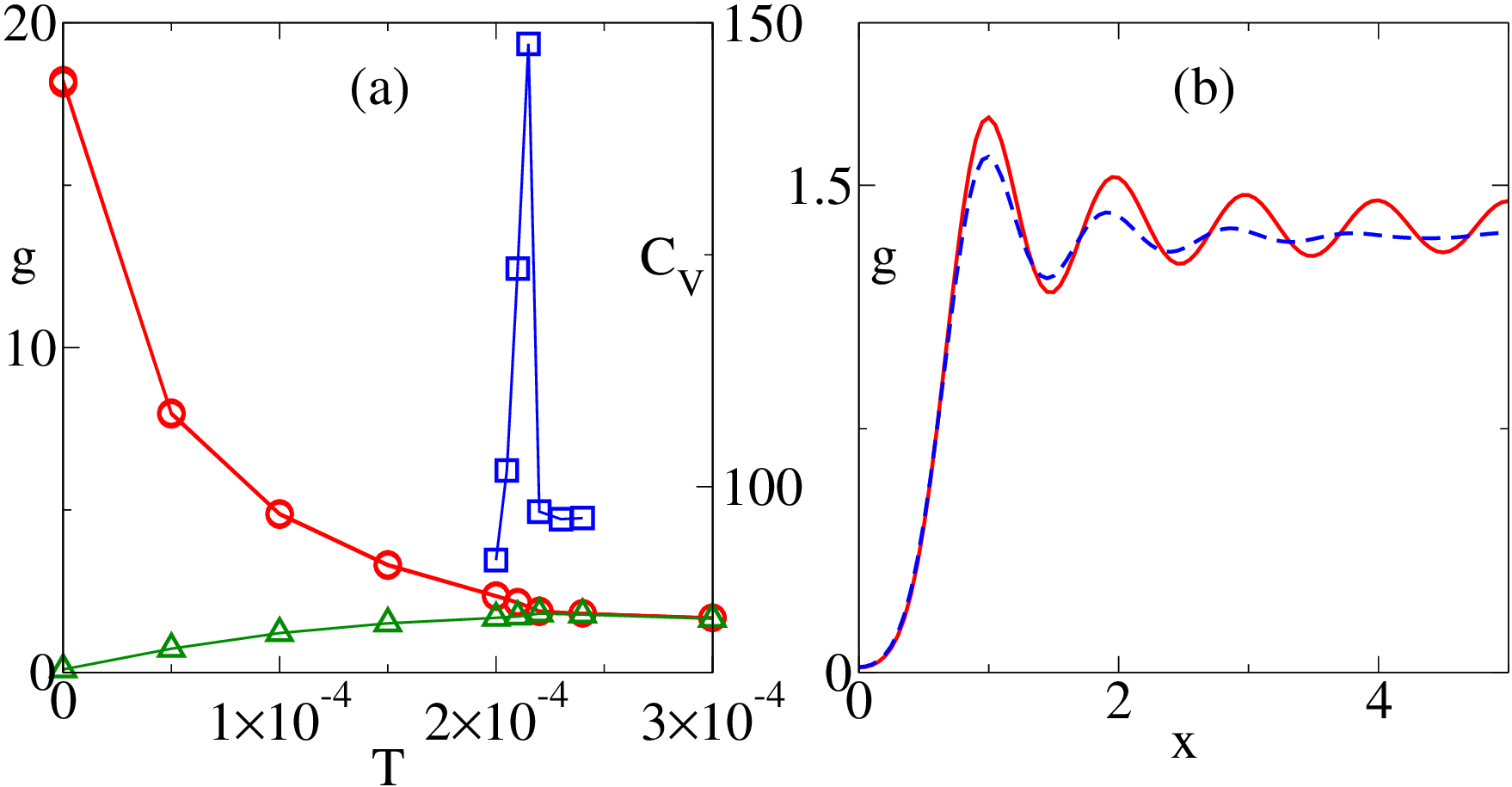}
\caption{ (color online)
(a) $C_V$ (blue squares), height of the correlation function first neighbor peaks (red circles) and background level (green triangles) versus temperature for $N_V=120$, $N_z=19$, $v=10^{-2}$.
(b) Amplitude of the in plane correlation function along the $x$ axis just below ($T=2.1 \times 10^{-4}$, red line) and above ($T=2.2 \times 10^{-4}$, dashed blue line) the transition.
}
\label{Cvcor}
\end{figure}
To elucidate the nature of the phase obtained once MBoG has disapeared, 
we compute in the $xy$ plane the pair correlation function $g(\textbf{r})=\langle\rho(\textbf{r'}+\textbf{r}) \rho(\textbf{r'})\rangle$ (Ref. 13), where $\rho(\textbf{r})$ is the vortex density, and the mean square displacement $B_1(t)=\langle [\textbf{r}(t)-\textbf{r}(0)]^2\rangle$ (Ref. 13), both at zero tilt. At low temperature, the correlation function displays a triangular lattice of peaks, in agreement with the expectation of a quasi ordered lattice for an elastic moving glass, while above a critical temperature we find circular rings and short range order, signature of a disordered isotropic phase. In order to compare the critical temperature obtained from the correlation function and the one obtained from $C_V$, we calculate the height of the peaks corresponding to the six first neighbors and the background level. As shown in Fig.\ \ref{Cvcor}a, the collapse of these two quantities --- which indicates the loss of the sixfold symmetry --- occurs at the same temperature $T_C$ as the peak in $C_V$. In Fig.\ \ref{Cvcor}b we plot the correlation function along the $x$ axis just below and above $T_C$, pointing out the change in the order range.
This picture is confirmed by the study of the mean square displacement. Below the critical temperature $B_1(t)$ is bounded, while above it grows linearly indicating a diffusive wandering as expected in a liquid. However, the diffusion coefficient is found to be a nonlinear function of the temperature, suggesting that the motion is more complex than a classic random walk.
To sum up, at a critical temperature $T_C$ we simultaneously see the vanishing of both the DTME and the lattice order, while the vortex displacements go from bounded to diffusive. We conclude that the transition observed is the dynamical melting of the MBoG into a moving vortex liquid (MVL).
\begin{figure}[h]
\centering
\includegraphics[width=0.95\linewidth]{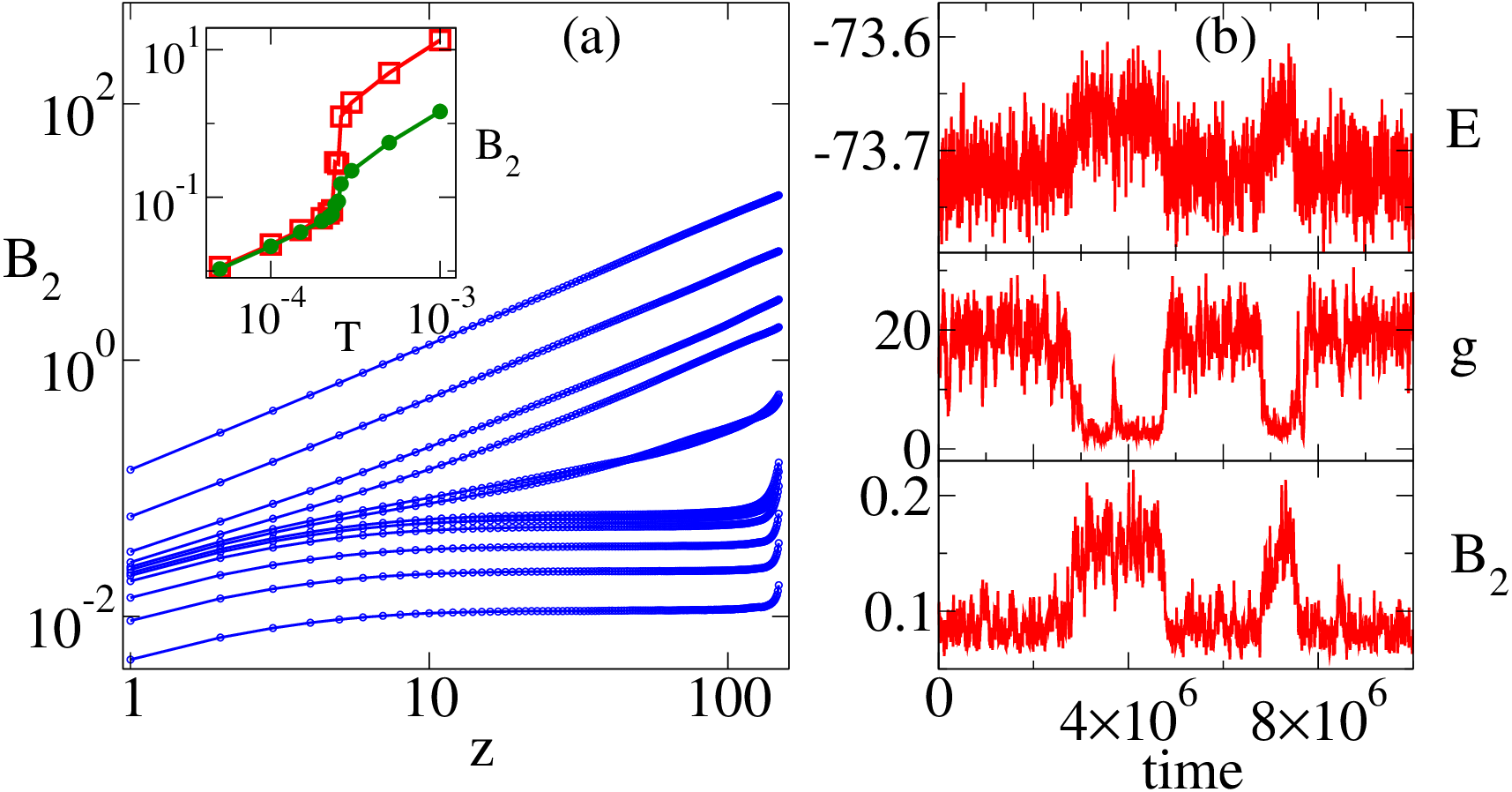}
\caption{ (color online)
(a) $B_2(z_{ij})=\langle (y_i-y_j)^2 \rangle$ ($z_{ij}$ is expressed in units of layer spacing $s$) for $T$ from $10^{-4}$ to $10^{-3}$, $N_V=30$, $N_z=149$, $v=10^{-2}$. Insert : $B_2(z_{ij})$ versus $T$  for $z_{ij}=10 s$ (green circles) and $z_{ij}=100 s$ (red squares).
(b) From top to bottom : total energy, amplitude of the correlation function at the first neighbor peaks and $B_2(L/2)$ versus time close to melting for $N_V=120$, $N_z=29$, $v=10^{-2}$.
}
\label{corz}
\end{figure}
Finally, we take a look at the correlations along the vortex lines in order to highlight the influence of the effective pinning potential $V$.
We plot in Fig.\ \ref{corz}a the mean square displacement in the transverse direction within the lines $B_2(z_{ij})=\langle (y_i-y_j)^2 \rangle$ (Ref. 13,14) where $i$ and $j$ are two pancakes belonging to the same flux line and $z_{ij}$ is the distance between the two layers they belong to.
At low temperature $B_2$ is bounded, each line being pinned in a minimum of $V$.
This pinning effect weakens as temperature is increased, and eventually vanishes at $T = T_C$. The insert of Fig.\ \ref{corz}a clearly displays the jump in $B_2(T)$ associated with this transition. The high temperature behavior can be understood by assuming that only thermal fluctuations and line cohesion are relevant : the length of each bond between two neighboring pancakes is then an independent random variable, and the line configuration is analog to a random walk in which $z$ would be the time, leading to the observed linear growth. We conclude that the loss of localization along the $z$ axis is a manifestation of the disappearance of the effective pinning potential. 
To be sure that this transition and the in-plane melting are two aspects of the same phenomenon, we monitor the time evolution of the quantities indicating the transition. 
 Because of finite size, close enough to the transition the system hesitates between different phases, and continuously switches from one to another.
We can check in Fig.\ \ref{corz}b that the three indicators of the transition plotted versus time (total energy, amplitude of correlation function at first neighbors peaks and $B_2(L/2)$) oscillate between two states, which we identify respectively with MBoG (lower energy) and MVL (highest energy), and that in-plane and out-of-plane quantities jump simultaneously when the system goes back and forth from one to the other. The effective pinning potential thus persists in MBoG whatever the temperature, and only vanishes when melting occurs.

To conclude, we find evidence of the MBoG phase at finite temperature, exhibiting DTME below a critical transverse field. It is stable in a wide range of temperature and velocity, respectively up to the melting temperature and to the highest velocities we are able to simulate.
Weaker pinning, which should lower the critical velocity $v_c$, has also been studied, but no evidence of the CMG phase has been found in that case either. A reduced density of pinning centers should also lower $v_c$, and could be a direction to further investigate the existence of CMG.
%
We also predict the existence of an effective transverse static tin roof pinning potential in the MBoG phase. Since the CMG doesn't exhibit DTME, we expect this potential to vanish in the CMG as it does in the MVL. 

We are grateful to Pierre Le Doussal and Kay J\"org Wiese for helpful and stimulating discussions.



\begin{thebibliography}{10}
\bibitem{linemelting}
  $B_2(z)$ remains finite even at $T>T_C$ because the layers remain coupled in the range of temperature we study.
\end{thebibliography}
\end{document}